# Hard spheres model of the atom

Roumen Tsekov
Department of Physical Chemistry, University of Sofia, 1164 Sofia, Bulgaria

The finite size effect of electron and nucleus is accounted for in the model of atom. Due to their hard sphere repulsion the energy of the 1s orbital decreases and the corrections amount up to 8% in Uranium. Several models for boundary conditions on the atomic nucleus surface are discussed as well.

A century ago Bohr has proposed a model of atom [1], which revolutionized physics and chemistry. It was refined later in quantum mechanics via the Schrödinger equation. The Bohr model describes nucleus and electrons as charged mass points but in fact they possess their own sizes. Some attempts are made in the middle of the previous century [2, 3] to estimate the effect of the finite size nucleus on the electron structure. The scope of the present paper is to reconsider again this problem and to get simple and transparent evaluations for the effect of the particle finite size on the most affected 1s electron for any element in the Periodic Table. The charges of electron and nucleus are considered to be located in their centers. Thus, the influence of the charge distribution [4] into the nucleus and electron is out of the focus of this paper.

For the sake of simplicity we consider here the 1s orbital of a Hydrogen-like atom. According to quantum mechanics the 1s electron wave function $\psi$ obeys the Schrödinger equation

$$-\frac{\hbar^2}{2m}\frac{1}{r^2}\frac{\partial}{\partial r}(r^2\frac{\partial}{\partial r}\psi) - \frac{Ze^2}{4\pi\varepsilon_0 r}\psi = E\psi \qquad (1)$$

where $Z$ is the atomic number. Introducing the Bohr radius $a_0 \equiv 4\pi\varepsilon_0\hbar^2/me^2$ and the dimensionless distance $\rho \equiv r/a_0$, Eq. (1) acquires the form

$$\frac{1}{\rho^2}\frac{\partial}{\partial\rho}(\rho^2\frac{\partial}{\partial\rho}\psi) + \frac{2Z}{\rho}\psi + \frac{2ma_0^2 E}{\hbar^2}\psi = 0 \qquad (2)$$

The standard solution of this equation is the first Slater-type orbital (STO) $\psi = \exp(-Z\rho)$ with the electron energy

$$E/mc^2 = -Z^2\alpha^2/2 \tag{3}$$

where $c$ is the speed of light and $\alpha = e^2/4\pi\varepsilon_0\hbar c = 1/137.036$ is the fine structure constant. The electron probability density of the 1s orbital equals to $f = 4Z^3\rho^2 \exp(-2Z\rho)$ and possesses a maximum at $\rho = 1/Z$, where the Bohr orbital lies.

As mentioned in the beginning, one of the shortcomings of the Bohr model is that the electron and nucleus are considered as mass points. Although their internal structures are not completely known, their finite sizes imply restriction on the electron wave function. The general solution of the Schrödinger equation (2) reads

$$\psi = U(1 - Z/\zeta, 2, 2\zeta\rho)\exp(-\zeta\rho) \tag{4}$$

where $U$ is the Kummer function and $\zeta \equiv \sqrt{-2ma_0^2 E/\hbar^2}$ is the effective charge of the nucleus. Since the size of the nucleus is of the order of femtometers, the corrections are relatively small and $\zeta$ is close to $Z$. Thus, expanding the Kummer function in series of $\zeta$ and keeping the linear term results in the following approximate wave function

$$\psi = \exp(-\zeta\rho) + \frac{\zeta - Z}{2Z^2\rho}\exp(-\zeta\rho) \tag{5}$$

As is seen, Eq. (5) is a superposition of the first and zeroth STOs.

Regarding the steric electron-nucleus interaction we propose the hard sphere model. In this case the wave function should vanish if the relative distance becomes $\rho_0 = (r_e + r_n)/a_0$. For the electron we employ the classical radius definition $r_e = \alpha^2 a_0 = 2.82$ fm, while the radius of the nucleus $r_n = r_e A^{1/3}/2$ scales with the third root of the mass number $A$. Thus, the minimal distance between the electron and nucleus equals to $\rho_0 = (1 + A^{1/3}/2)\alpha^2$. Substituting the wave function from Eq. (5) in the boundary condition $\psi(\rho_0) = 0$ results in an expression for the effective charge of the nucleus

$$\zeta = Z - 2Z^2\rho_0 = Z - (2 + A^{1/3})Z^2\alpha^2 \tag{6}$$

As is seen $\zeta < Z$ which corresponds to increase of the electron energy due to repulsion of the electron from the nucleus via the steric interaction. According to Eq. (6) the effective charge of the Hydrogen atom $^1_1H$ is $\zeta = 0.9998$, which is very slightly lower than $Z = 1$. However, in the atom of Uranium $^{238}_{92}U$ the effective charge equals to $\zeta = 88.303$, which corresponds to 8 % higher energy of the 1s electron. More precisely $\zeta$ can be calculated directly from the equation $U(1-Z/\zeta, 2, 2\zeta\rho_0) = 0$, which provides $\zeta = 88.893$ for Uranium (see Fig. 1).

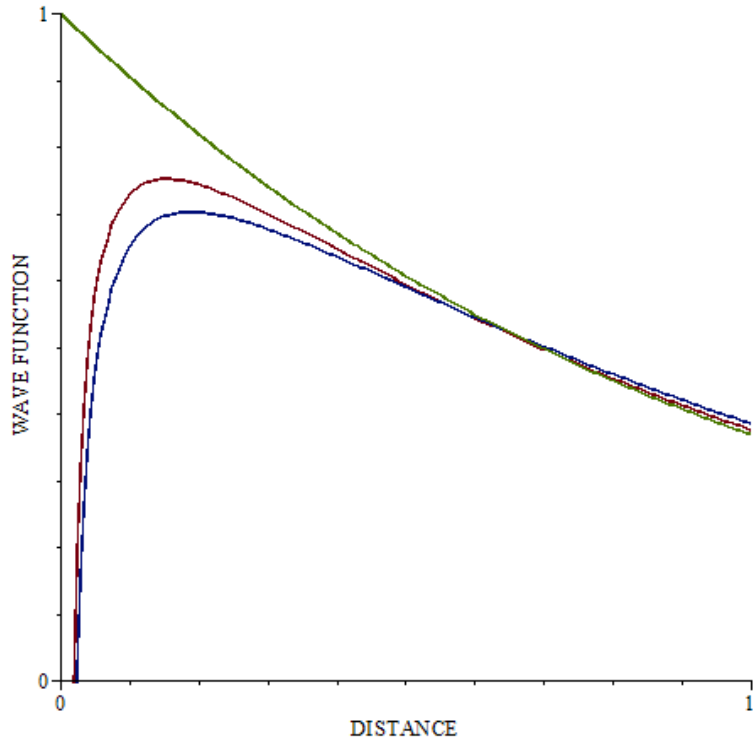

**Fig. 1** The wave function $\psi$ versus the distance $Z\rho$ for Uranium: the classical exponential expression (green), the exact solution (4) (red) and the approximation (7) (blue)

Substituting Eq. (6) in Eq. (5) yields to

$$\psi = (1 - \rho_0/\rho)\exp(-\zeta\rho) \qquad (7)$$

and the corresponding electron probability density $f = 4\zeta^3(\rho-\rho_0)^2 \exp[-2\zeta(\rho-\rho_0)]$ possesses a maximum at $\rho = 1/\zeta + \rho_0 = 1/Z + 3\rho_0$. Hence, the position of the Bohr orbital is also shifted.

The electron energy equals to $E/mc^2 = -\zeta^2\alpha^2/2$ and the electron is bounded weaker than in the Bohr model (3) since $\zeta < Z$. Introducing the effective charge from Eq. (6) yields

$$E/mc^2 \approx -Z^2\alpha^2/2 + (2+A^{1/3})Z^3\alpha^4 \tag{8}$$

One can see the positive correction here, which originates from the nucleus-electron steric repulsion. The corresponding force acting on the nucleus $F \equiv \partial E/\partial r_n = 2\hbar^2 Z^3/ma_0^3 = 0.165 Z^3$ µN is sufficiently large for heavy atoms. Surprisingly due to the relativistic expression used for the electron radius the correction in Eq. (8) is similar to the relativistic corrections to the Bohr model. Indeed, the Dirac relativistic quantum theory [5] results in the following expression

$$E/mc^2 = \sqrt{1-Z^2\alpha^2} - 1 \approx -Z^2\alpha^2/2 - Z^4\alpha^4/8 \tag{9}$$

but the sign of the relativistic Sommerfeld correction [6] is negative. Hence, the relativity itself makes the atom more stable. However, the dependence of these two corrections on the atomic number is different. The finite size correction is larger than the relativistic one for the light elements up to the f-block, while for heavier chemical elements the relativistic correction dominates. The combined effect leads to $\zeta = Z - (2 + A^{1/3} - Z/8)Z^2\alpha^2$. Thus, Lanthanum $^{139}_{57}La$ should obey the classical Bohr model since the finite size and relativistic corrections cancel each other.

In the present paper the nucleus is modelled as a sphere impenetrable for the electron. Let us reconsider the standard boundary condition $\psi(\rho_0) = 0$ for the simpler case of a quantum particle into infinitely deep potential well. In this case the Schrödinger equation

$$-\frac{\hbar^2}{2m}\frac{d^2\psi}{dx^2} = E\psi \tag{10}$$

possesses the following general solution

$$\psi = C_1 \sin(\sqrt{2mE}x/\hbar) + C_2 \cos(\sqrt{2mE}x/\hbar) \tag{11}$$

The infinite potential energy on the well borders requires application of the Dirichlet boundary conditions $\psi(0) = \psi(L) = 0$, which results in wave functions $\psi_n = \sqrt{2/L}\sin(\sqrt{2mE_n}x/\hbar)$ with a discrete energy spectrum $E_n = (n\pi\hbar)^2/2mL^2$, $n = 1,2,3,...$ Usually, the same solution is considered to describe a quantum particle in a box. However, in this case the quantum particle undergoes elastic collisions at the box walls, which are characterized by change of the direction of the particle velocity and return to the state before the collision. Hence, according to the Rolle theorem the derivative of the wave function must vanish at the box borders [7]. Applying the relevant Neumann boundary conditions $\psi'(0) = \psi'(L) = 0$, the general wave function (11) reduces now to $\psi_n = \sqrt{2/L}\cos(\sqrt{2mE_n}x/\hbar)$, but the energy spectrum $E_n$ remains the same. Surprisingly, the uniform wave function $\psi_0 = 1/\sqrt{L}$ is also a solution of Eq. (10) and the corresponding energy $E_0 = 0$ is zero. If we apply the Neumann boundary condition $\psi'(\rho_0) = 0$ to the wave function (5) the effective charge equals to $\zeta = Z - 2Z^3\rho_0^2$ and the correction is smaller now than that in Eq. (6). An interesting model is that of mixed boundary conditions. In the cases of $\psi(0) = \psi'(L) = 0$ or $\psi'(0) = \psi(L) = 0$ the wave function (11) acquires the forms, respectively,

$$\psi_n = \sqrt{2/L}\sin(\sqrt{2mE_n}x/\hbar) \qquad \psi_n = \sqrt{2/L}\cos(\sqrt{2mE_n}x/\hbar) \qquad (12)$$

However, the energy spectrum $E_n = (n\pi\hbar)^2/2mL^2$ with quantum numbers $n = 1/2, 3/2, 5/2,...$ differs from the classical expression above. More complex models involve the Robin boundary conditions as well [8]. Thus, the boundary condition for the electron on the nucleus surface is still an open question. One could imagine, for instance, partially penetrable nucleus, partially absorbed electron or even K-electron capture [9]. These effects can be described via the Robin boundary condition $\psi'(\rho_0) = \beta\psi(\rho_0)$, which includes the Dirichlet and Neumann boundary conditions at the limiting cases $\beta \to \infty$ and $\beta \to 0$, respectively. In this case the effective charge $\zeta = Z - 2Z^2\rho_0^2(Z+\beta)/(1+\beta\rho_0)$ possesses a singularity at $\beta \to -1/\rho_0$. If $\beta = -Z$ than $\zeta = Z$ and the electron wave function (4) reduces to $\psi = \exp(-Z\rho)$. Thus, the electron is freely tunneling through the nucleus, which seems transparent to it.

A way to describe the finite size effect without boundary conditions is to introduce a relevant repulsive potential in the system Hamiltonian. Since the energy from Eq. (8) depends linearly on $Z\rho_0$ we propose the following extension of Eq. (2)

$$\frac{1}{\rho^2}\frac{\partial}{\partial\rho}(\rho^2\frac{\partial}{\partial\rho}\psi)+\frac{2Z}{\rho}\psi-\frac{2Z\rho_0}{\rho^2}\psi-\zeta^2\psi=0 \tag{13}$$

where the nucleus-electron steric interaction is described by the term $Z\rho_0/\rho^2$. At large distances the Coulomb attraction prevails over the repulsion, while at short distance the repulsion is dominant. Thus, there is a minimum of the potential energy at $2\rho_0$. The solution of Eq. (13) is a fractional STO $\psi=\rho^{2Z\rho_0}\exp(-\zeta\rho)$, which possesses a maximum similar to Fig. 1. The effective charge of the nucleus is given by Eq. (6) again. Therefore, the finite size effect of the nucleus and electron can be well accounted for by an effective Coulomb potential

$$V_{ne}=(r_0-r)Ze^2/4\pi\varepsilon_0 r^2 \tag{14}$$

Finally, the effect of collisions between electrons is also a challenge to explore [10]. For instance, the core electrons do not only screen the electrostatic interaction between the valence electrons with the nucleus but they could increase the effective size of the latter as well due to electron-electron collisions. To examine the effect of a second electron on the 1s orbital let us consider a Helium-like atom, the energy of which is given by the expression

$$E/mc^2=(4\pi\alpha)^2\int_{\rho_0}^{\infty}\int_{\rho_0}^{\infty}[\frac{1}{2}(\frac{\partial\psi}{\partial\rho_1})^2+\frac{1}{2}(\frac{\partial\psi}{\partial\rho_2})^2-(\frac{Z}{\rho_1}+\frac{Z}{\rho_2}-\frac{1}{\rho_{12}})\psi^2]\rho_1^2\rho_2^2 d\rho_1 d\rho_2 \tag{15}$$

Since it is not possible to solve analytically the Schrödinger equation in this case, one can employ the variation method. The product of Hydrogen-like functions $\psi=\psi_1\psi_2$ is usually chosen as an approximate wave function, where the single electron wave functions are given by Eq. (7) now. However, the product state does not account for the electron-electron correlation. Since the two electrons are equally charged they repel strongly each other and perhaps their real position is when the two electrons are always diametrically opposite in the atom. Taking into account this correlation correction we replace the inter-electronic distance $\rho_{12}$ by its maximal value $\rho_1+\rho_2$. In this case the integration of Eq. (15) is possible and the energy of the atom acquires the analytical form

$$E/mc^2 = -(2Z-\zeta-2/5-4Z\zeta\rho_0+2\zeta\rho_0/5)\zeta\alpha^2 \qquad (16)$$

Let us consider first the traditional case of point charges. Setting $\rho_0 = 0$ Eq. (16) leads to the expression $E_0/mc^2 = -(2Z-\zeta-2/5)\zeta\alpha^2$. Minimizing it in respect to the effective charge results in $\zeta_0 = Z-1/5$ and $E_0/mc^2 = -\zeta_0^2\alpha^2$. Thus, the electron-electron repulsion reduces the nucleus charge by $1/5$. Introducing now the linear model $\zeta = \zeta_0 - C\rho_0$ in Eq. (16) and minimizing the result in respect to the still unspecified constant $C$ yields

$$\zeta = \zeta_0(1-2\zeta_0\rho_0-\rho_0/5) \qquad (17)$$

One can easily prove that the corresponding energy $E/mc^2 = -\zeta^2\alpha^2$ is correctly expressed by the effective charge. Equation (17) shows that the electron-electron interaction increases the effect of the finite size nucleus apart from the usual reduction of $\zeta_0$. Thus, for Uranium the effective charge decreases further to $\zeta = 88.115$ according to Eq. (17). We obtained the same results via the effective potential (14), when the electron-electron interaction is also properly adjusted to $V_{ee} = (r_1+r_2-2r_0)e^2/4\pi\varepsilon_0(r_1+r_2)^2$.